\documentclass[showpacs,twocolumn,prb,eqsecnumbers]{revtex4}
\usepackage{latexsym}
\usepackage{graphicx}

\begin{document}
\title{Exploring of point-contact spectra of Ba$_{1-x}$Na$_{x}$Fe$_{2}$As$_{2}$
in the normal and superconducting state }

\author{Yu. G. Naidyuk, O. E. Kvitnitskaya}

\affiliation{B.\,Verkin Institute for Low Temperature Physics and Engineering, National Academy of Sciences of Ukraine,
47 Lenin Ave., 61103, Kharkiv, Ukraine}

\author{S. Aswartham, G. Fuchs, K. Nenkov, S. Wurmehl }

\affiliation{Leibniz-Institut f\"{u}r Festk\"{o}rper- und Werkstoffforschung Dresden e.V., Postfach 270116, D-01171 Dresden, Germany}

\begin{abstract}

We present study of derivatives of current-voltage $I(V)$ characteristics of point-contacts (PCs)
based on Ba$_{1-x}$Na$_{x}$Fe$_{2}$As$_{2}$ ($x$=0.25) in the normal and
superconducting state. The detailed analysis of \textit{dV/dI(V)} data (also given in Appendix A)
shows that the thermal regime, when temperature
increases with a voltage at a rate of about 1.8 K/mV, is realized in the
investigated PCs at least at high biases $V$ above the superconducting (SC) gap
$\Delta $. In this case, specific resistivity $\rho (T)$ in PC
core is responsible for a peculiar \textit{dV/dI(V)} behavior, while a pronounced
asymmetry of \textit{dV/dI(V)} is caused by large value
of thermopower in this material. A reproducible zero-bias
minima detected on \textit{dV/dI(V)} at low biases in the range $\pm$(6--9)\,mV well below
the SC critical temperature $T_{c}$ could be connected with the manifestation
of the SC gap $\Delta $. Evaluation of these Andreev-reflection-like
structures on \textit{dV/dI(V)} points out to the preferred value of 2$\Delta
$/k$_{\rm B}T_{c}\approx 6$. The expected second gap features on
\textit{dV/dI(V)} are hard to resolve unambiguously, likely due to impurity scattering,
spatial inhomogeneity and transition to the mentioned thermal regime as the
bias further increases. Suggestions are made how to separate spectroscopic
features in \textit{dV/dI(V)} from those caused by the thermal regime.

\end{abstract}


\pacs{73.40.Jn, 74.70.Dd, 74.45.+c}

\maketitle

\section{INTRODUCTION}

Yanson's point-contact (PC) spectroscopy is the time-proved technique to
investigate electron-phonon (quasiparticle) interactions in solids \cite{YansonZETF, Atlas}.
PC spectroscopy was successfully applied to a large number of simple metals
and more complex conducting systems, thus it became a world-wide recognized
method in the solid state physics \cite{PCSbook}. One of the related (sister) branches
of PC spectroscopy to investigate the superconducting (SC) state is PC
Andreev-reflection (PCAR) spectroscopy. The latter is the well-reputed tool
to get SC parameters from the PC differential resistance/conductance.
Combining these two approaches gives the possibility to receive
simultaneously information about the normal as well as the SC state
properties of the matter under study. During the last years PCAR
spectroscopy is widely used to investigate SC state in newly discovered
iron-based superconductors (see, e.g., Ref. \cite{Daghero}). The most attention here
is paid to the study of the SC gap(s) and its temperature and directional
dependence. However, study of the normal state behavior of these materials
by Yanson's PC spectroscopy attracted less attention. One of the reasons is
that Yanson's PC spectroscopy of quasiparticle excitations is carried out at
energies usually an order of amplitude larger than the SC gap energy. In
this case shortening of the inelastic mean free path of electrons can
result to violation of spectroscopy because of transition to the thermal
regime when the temperature rises in the PC core by increasing a bias
voltage \cite{Verkin}. Nevertheless, study of the normal state of iron-based
superconductors by Yanson's PCs can provide helpful information as to the
current regime in PCs as well as to the quasiparticle excitations. In this
paper we have carried out comprehensive PC measurements in the normal and in the
SC state on single crystals of Ba$_{1-x}$Na$_{x}$Fe$_{2}$As$_{2}$ - typical
member of the mentioned iron family with [122] structure.

\section{EXPERIMENTAL DETAILS}

Single crystals of Ba$_{1-x}$Na$_{x}$Fe$_{2}$As$_{2}$ were grown using a
self-flux high temperature solution growth technique. Substitution of Ba by
Na leads to the suppression of spin-density wave ordering and induces
superconductivity up to 34\,K at optimal doping for $x\approx$ 0.4\cite{Aswartham}. We studied
samples with the intermediate sodium concentration $x\approx$ 0.25. For a single
crystals with $x\approx$ 0.25 the resistivity shows a kink at $T=$ 117\,K (see Ref. \cite{Aswartham} and
inset in Fig.\,\ref{fig9}), typically found in hole-doped [122]-type compounds (see,
e.g., Ref. \cite{Shen}). This kink is an indication of the structural and magnetic
transition. Thus, the sample with $x\approx$ 0.25 shows both an antiferromagnetic
and a SC transition what is typically present in the underdoped [122]-type
compounds. The SC transition occurs here at about 10\,K \cite{Aswartham}.

\begin{figure}[htbp]
\centerline{\includegraphics[width=9cm,angle=0]{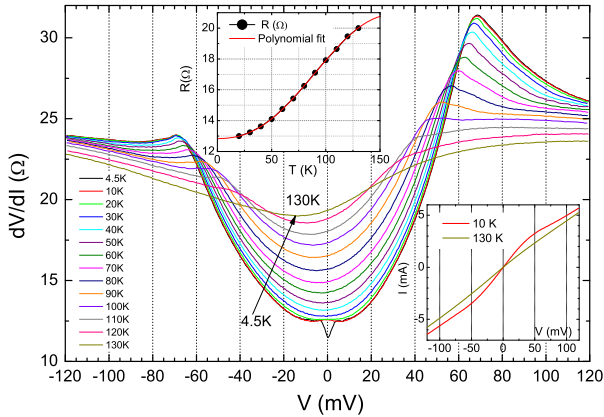}}
\caption{(Color online) \textit{dV/dI(V)} measured for the Ba$_{0.75}$Na$_{0.25}$Fe$_{2}$As$_{2}$--Cu PC
with $R\approx 13\,\Omega$ at various temperatures. Upper inset:
Temperature dependence of resistance $R(T)=dV/dI(V=0,T)$ of this PC (dots) measured
at $V=0$ along with a polynomial approximation (solid curve). Bottom inset: examples of
$I(V)$ characteristics for PC from the main panel measured above $T_c$ at 10\,K and 130\,K.}
\label{fig1}
\end{figure}

We have chosen Ba$_{1-x}$Na$_{x}$Fe$_{2}$As$_{2}$ single crystals with $x\approx$ 0.25
and relative low $T_{c}\approx$10\,K to search for electron-quasiparticle
interaction features in the normal state, that is above $T_{c}$, by Yanson's
PC spectroscopy. In this case thermal resolution (smearing) of PC spectra \cite{Atlas,PCSbook} of about
5.44k$_{\rm B}T \ge $ 5\,meV is still not high to prevent spectroscopic
investigation. On the other hand, measuring above $T_{c}$ would prevent
appearance of SC features, which can mask quasiparticle excitations
(especially in the low energy region) or may be erroneously taken as the
latter. We should mention that the investigated single crystals were
nonhomogeneous (at least at the surface), probably, due to volatility of Na
component and showed a significant variation of the local $T_{c}$ of PC as it
will be shown below.

The PCs were established \textit{in situ }by touching a sharpened Ag or Cu thin wire
(${\o}\approx$ 0.2--0.3\,mm) to the cleaved (at room temperature)
surface or to an edge of the sample. Thus, we have measured heterocontacts
between normal metal and the title compound. The differential resistance
\textit{dV/dI(V)}$\equiv R(V)$ and $dV^2/dI^{2}(V)$ of $I(V)$ characteristic of PC were recorded by sweeping
the \textit{dc} current $I$ on which a small \textit{ac} current $i$ was superimposed using the standard
lock-in technique. The measurements were performed in the temperature range
from 3\,K up to 130\,K in some cases.

\section{RESULTS AND DISCUSSION}

Spectroscopic information including the determination of the SC energy gap
can be derived from PC measurements only if the investigated PCs are in the
ballistic or in the diffusive regime of the current flow \cite{PCSbook}, where
electrons can gain an excess energy up to e$V$ and heating effects are
negligible. Thus, the contact size $d$ should be smaller than the inelastic
electron mean free path. To shrink a variation of the SC gap on the scale of
the PC size, the PC diameter $d$ should also be less than the SC coherence
length $\xi$, which amounts to only 2\,nm in the isostructural system
Ba$_{1-x}$K$_{x}$Fe$_{2}$As$_{2\,}$\cite{Wray}.

\begin{figure}[htbp]
\centerline{\includegraphics[width=9cm,angle=0]{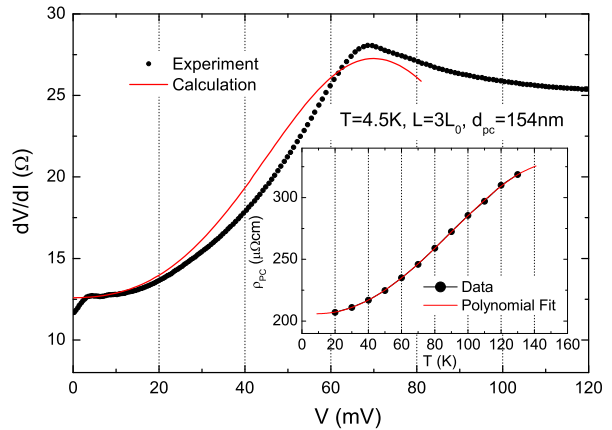}}
\caption{(Color online) Symmetrised \textit{dV/dI(V) }of the PC (points) from Fig.\,\ref{fig1} measured at T$=$4.5\,K along
with a calculated \textit{dV/dI(V)} according to Eqs.\,(\ref{eq2}),(\ref{eq3}) (solid line) with $d$=154\,nm. To fit the position of
the maxima we used an enhanced Lorenz number $L=3$\textit{L}$_{0}$ in the calculation.
Inset: $\rho_{PC}(T)$ used for the calculations, which mimics the
$R(T)$ behavior from Fig.\,\ref{fig1} (see inset).}
\label{fig2}
\end{figure}

The PC diameter $d$ can be estimated from its resistance $R_{PC}$ according to
Wexler's formula \cite{PCSbook}:

\begin{equation}
\label{eq1}
R_{PC}\approx 16\rho l/3\pi d^{2}+\rho /2d.
\end{equation}

Here we neglect the contribution to the second term related to the
resistivity of a clean normal metal (needle) and suppose geometrically equal
parts occupied by each metal in our case of heterocontact. Utilizing the
Drude free electron model, we estimated $\rho l \approx 1.3 \cdot
10^{4}n^{-2/3}\approx 3.8\cdot 10^{-11}\Omega \cdot
$cm$^{2}$ using the electron density $n \approx 6.2\cdot 10^{21\,
}$cm$^{-3}$ obtained from the Hall coefficient $R_{H}\approx 4\cdot
10^{-3}$cm$^{3}$/C for Ba$_{1-x}$K$_{x}$Fe$_{2}$As$_{2}$ with $x=$0.4 \cite{Aswartham}. The elastic electronic
mean-free path in Ba$_{0.75}$Na$_{0.25}$Fe$_{2}$As$_{2}$ is estimated to be
about $l=\rho l/\rho_{0}\sim $10\,nm using the residual
resistivity of the investigated samples $\rho_{0} \approx $ 40\,$\mu
\Omega $cm \cite{Aswartham}. From Eq.\,(\ref{eq1}), the PC diameter $d$ is estimated to be between
14 and 60 nm using typical PC resistance between 5 and 50 $\Omega $. Thus,
the necessary condition of the smallness of the PC size $d$ compared to $l$ and
$\xi$ is not fulfilled for the investigated PCs. Therefore, a distribution
of SC gap is expected in the PC region and the current flow is at least
diffusive. The diffusive regime itself does not prevent to get spectral
information, however it favors the transition to the thermal regime with a
bias rise. So, the observation of characteristic AR features in the
\textit{dV/dI(V) }curves in the SC state is important to prove the spectral regime. Another
criteria of realization of the spectral regime in PCs is the observation of
phonon (quasiparticle) structures in $d^{2}$\textit{V/dI}$^{2}(V)$ curves in the normal
state.
\begin{figure}[htbp]
\centerline{\includegraphics[width=9cm,angle=0]{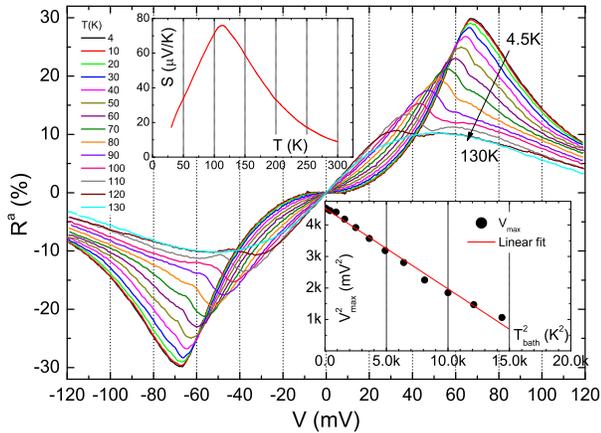}}
\caption{(Color online) Calculated antisymmetric part
$R^{a}(\%)=100$\textit{(R(V \textgreater }0)$-$\textit{R(V \textless }0))$/$2$R(V=0)$
of PC from Fig.\,\ref{fig1} at different temperatures, here $R=dV/dI$. Right inset: Position
of the maximum of $R^{a}$ vs temperature in quadratic coordinates. Left
inset: Temperature dependence of the thermopower $S$ of
Ba$_{0.7}$K$_{0.3}$Fe$_{2}$As$_{2}$ \cite{Yan}.}
\label{fig3}
\end{figure}

We have measured hundreds of \textit{dV/dI(V)} dependences of
Ba$_{0.75}$Na$_{0.25}$Fe$_{2}$As$_{2}$ -- Cu(or Ag) PCs.
In general, the \textit{dV/dI(V)} spectra do not show any principal difference
while being measured by attaching the needle to the cleaved surface or to an
edge of the samples. In both cases the shape of \textit{dV/dI(V)} variates from one PC to another PC.
This is because PC is created by chance and its microscopic structure is caused
by influence of stress, surface defects, oxides, impurities etc.
Therefore, we analyzed \textit{dV/dI(V)} to establish reproducible features in the PC spectra reflecting bulk
properties.

Taking into account the short characteristic electronic lengths in the
iron-based superconductors, as estimated above, and the steep resistivity
rise with temperature \cite{Aswartham}, one should distinguish carefully the
spectroscopic features from those caused by thermal effects.

\begin{figure}[htbp]
\centerline{\includegraphics[width=9cm,angle=0]{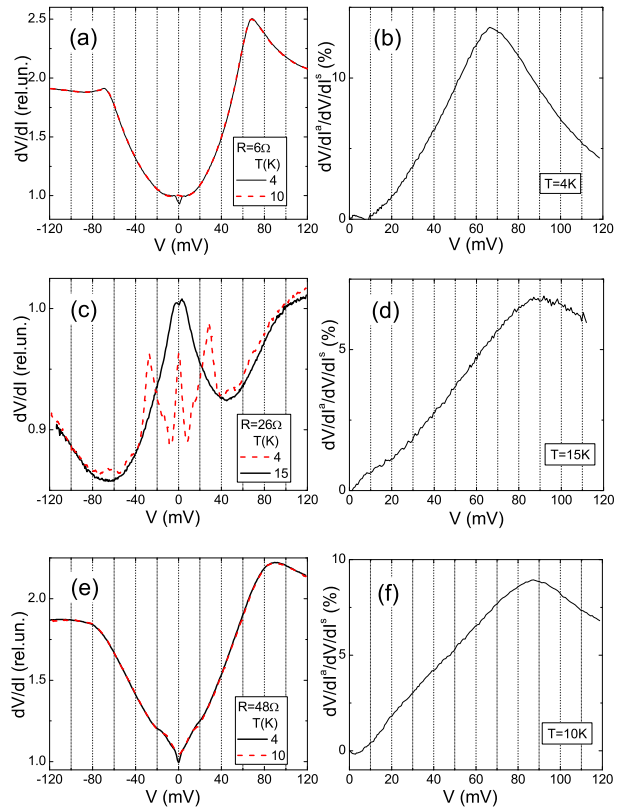}}
\caption{ (Color online) Left panels: \textit{dV/dI }curves for several
Ba$_{0.75}$Na$_{0.25}$Fe$_{2}$As$_{2}$--Cu contacts for a broad bias
range measured below and above $T_{c}$. Right panels: reduced antisymmetric part
\textit{dV/dI}$^{a}$\textit{(V)/dV/dI}$^{s}(V)$ for the corresponding curves from the left
panel. Here \textit{dV/dI}$^{s}(V)=$\textit{(dV/dI(V\textgreater }0)$+$\textit{ dV/dI(V\textless }0)$/$2$R(V=0)$ 
is symmetric part of \textit{dV/dI(V).}}
\label{fig4}
\end{figure}

Characteristic \textit{dV/dI(V)} curves measured in a wide voltage and temperature range
are shown in Fig.\,\ref{fig1}. Here, conspicuous maxima at about $\pm$70\,mV and pronounced
\textit{dV/dI(V)} asymmetry are seen. The maxima become broader and their position
shifts to lower energies as the temperature rises. Besides, the \textit{dV/dI(V)} at 4.5\,K displays a
small zero-bias minimum, which is of SC origin because it vanishes at
about 10\,K, what coincides with the $T_{c}$ of the bulk sample for $x=$\,0.25.
With temperature increase, the intensity of the main
parabolic-like minimum reduces and the side maxima move to lower voltages as
the temperature increases. In the inset, the temperature dependence of the
PC resistance $R(T)$ (defined as $R=dV/dI(V=0)$ is shown. Differentiating of
formula (\ref{eq1}) with respect to temperature, we find $d =$ ($d\rho
$\textit{/dT})/2(\textit{dR/dT}) and, thus, the PC size can be determined by measuring the temperature
dependence of its resistance. Taking the mentioned derivatives for several
temperatures between 20 and 80\,K, an average diameter of about 80\,nm is
obtained. Using this procedure to determine the PC diameter, the real
resistivity in the PC core can be derived.
The residual resistivity of the PC turned out to be about $\rho_{0}^{PC} \approx $
200\,$\mu \Omega $cm (calculated using $d$=80\,nm and Eq.\,(\ref{eq1}), where
a small contribution of the first term is neglected), what is 5 times larger than the bulk value!
Accordingly, the elastic mean free path of electrons in PC will be 5 times
smaller and amounts to only a few nanometers. All these estimations exclude
the ballistic regime and promote the realization of the thermal regime of
the current flow with a voltage rise. This point will be checked further by
analyzing the measured \textit{dV/dI(V)} in the thermal regime.

\section{Analyzing in the thermal regime }

In the case of thermal regime, when the contact size $d$ becomes much larger
than the inelastic electron mean-free path, the temperature inside PC
increases with the bias voltage according to the well-known relation \cite{PCSbook,Verkin}:

\begin{equation}
\label{eq2} T_{PC}^{2}=T_{bath}^{2}+V^{2}/4L_{0},
\end{equation}

where $T_{PC}$ is the temperature in the PC core, $T_{bath}$ is the
temperature of the bath, and $L_{0}$ is the Lorentz number
($L_{0}=2.45 \cdot 10^{-8\,}$V$^{2}$/K$^{2})$.

Using Kulik's thermal model \cite{Verkin,Kulik1}, the $I(V)$ characteristics of the PC and its
derivative can be calculated from the temperature dependence of the
resistivity $\rho (T)$ of the material under study:

\begin{equation}
\label{eq3}
I(V)=Vd\int\limits_0^1 \frac{dx}{\rho (T_{PC} \sqrt{(1-x^2)})}.
\end{equation}

Applying this formula, we have calculated \textit{dV/dI(V) }in the thermal regime using for
the temperature dependence of the resistivity  $\rho_{PC}(T)$ in PC a curve similar
to $R(T)$ as shows the inset of Fig.\,\ref{fig1} with the residual
resistivity $\rho_{0}^{PC} \approx $ 200\,$\mu \Omega$cm 
estimated above. The obtained result shown in Fig.\,\ref{fig2} demonstrates the good
qualitative and quantitative correlation with experimental data.
Here we should note that the calculations result in diameter twice as large as $d$=80\,nm
determined above from the temperature dependencies. The reason is that Eq.(3) is derived for homocontact.
We suppose that factor 2 must be included in the right part of Eq.(3) in the case of symmetric geometrically
heterocontact. Therefore $d$=80\,nm derived above should be compared with $d$/2=77\,nm from Fig.2.
Thus, the considered \textit{dV/dI(V)} curves correspond to the thermal regime of the current flow. A detailed
analysis of the calculation of \textit{dV/dI(V)} in the thermal regime using bulk $\rho (T)$ data
and the evolution of \textit{dV/dI(V)} for several temperatures is given below in Appendix A.

One of the pronounced features of the \textit{dV/dI(V)} characteristics is their asymmetry.
All measured \textit{dV/dI(V)} curves are highly asymmetric having larger \textit{dV/dI(V)} values for the
positive bias. Fig.\,\ref{fig3} presents the antisymmetric part $R^a(V)$ of
$dV/dI(V)$ for the PC from Fig.\,\ref{fig1}. The
calculated $R^{a}(V)$ demonstrates the pronounced maximum at about 70\,mV which
becomes suppressed and shifts to the lower voltages at increasing temperature.
Qualitatively, the shape of $R^{a}(V)$ curves at low temperature corresponds
well to the temperature dependence of the thermopower $S(T)$ measured for the
isostructural compound Ba$_{0.7}$K$_{0.3}$Fe$_{2}$As$_{2}$ \cite{Yan} (see left
inset in Fig.\,\ref{fig3}). Additionally, the sign of the asymmetry is opposite to
that found in SmFeAsO$_{1-x}$F$_{x}$ PCs \cite{Naidyuk1}, where the thermopower is
also large, but negative. Besides, the temperature dependence of the maximum
position in $R^{a}(V)$ follows a quadratic law (see right inset in Fig.\,\ref{fig3})
which is in accord with Eq.\,(\ref{eq2}). All this confirms that the investigated
PCs are in the thermal regime of the current flow, at least at high biases
in the region of the maximum. In the case of heterocontact, $R^{a}(V)$
resembles the thermoelectric power of the investigated sample \cite{Naidyuk2}. Also the
maximum of the normalized $R^{a}(V)$ in Fig.\,\ref{fig3} is high (up to 30{\%}), what also
coincides with the exceptionally high $S$ as seen from Fig.\,\ref{fig3} (left inset).

The Lorentz number $L$ in the investigated PC estimated from the slope of the
$T_{bath}^{2}$ vs $V^{2}$ plot (see right inset in Fig.\,\ref{fig3}) was found to
deviate from the standard value being equal to about 2.5$L_{0}$ which is
close to the value we used for the calculation in Fig.\,\ref{fig2}. This is an
additional prove that the characteristic features of \textit{dV/dI(V)} -- maxima and
asymmetry can be described by the thermal model. An enhanced $L$ may be due to the additional
contribution of phonons to the thermal conductivity of PC through electrically
non-conductive osculant surfaces.

We collected \textit{dV/dI(V) }curves of different shape for several PCs in the left panels of
Fig.\,\ref{fig4}. Nevertheless, in all cases, the corresponding $R^{a}(V)$ data (right
panels) show similar behavior and exhibit a maximum between 70 and 90\,mV.
The similar shape of $R^{a}(V)$ for different types of PCs testifies that this
phenomenon is robust and that the asymmetry is mainly determined by the
properties of the less disturbed bulk material (see Appendix B) while the PC
itself plays here the role of a ``heater'' or "hot spot".

We suggest that the thermal regime is realized also for PCs of other
iron-based superconductors, which have similar high resistivity and
thermopower. Therefore, \textit{dV/dI(V)} data in this case provide information that comes
from $\rho (T)$ modified in PC core. The authors of Ref. \cite{Arham} reported, that they succeed to get
PCs on [122]-type compounds free from heating. Their claim is based on the
deviation of the behavior of the measured $dV/dI(V)=R(V)$ from that of $\rho (T)$.
Thus, they state (see paragraph  before summary in \cite{Arham}), that the evidence of the non-thermal
regime is "no agreement of bulk resistivity with $dV/dI$". At the same time, according to Eq.(1), $R(T)$
must behave similar to $\rho (T)$ in any case independently on the current regime, what we also
demonstrate here (compare $R(T)$ and $\rho (T)$ in the insets of Fig.\,\ref{fig1} and Fig.\,\ref{fig10}).
Therefore, the interpretation of the PC data in \cite{Arham} should be not taken too
literally.
\begin{figure}[htbp]
\centerline{\includegraphics[width=9cm,angle=0]{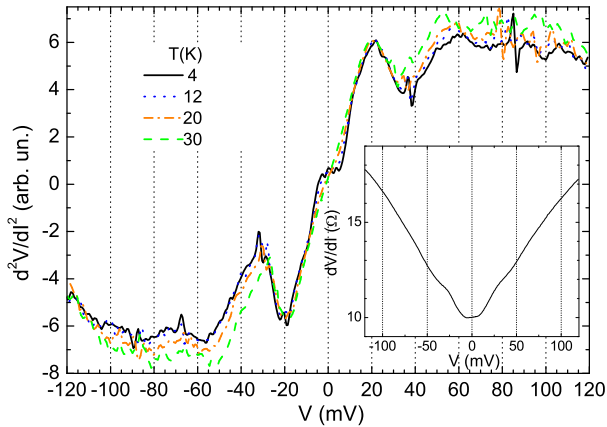}}
\caption{(Color online) $d^2V/dI^{2}(V)$ of the Ba$_{0.75}$Na$_{0.25}$Fe$_2$As$_2$--Cu PC
with $R=10\,\Omega $ measured at several temperatures. Inset: $dV/dI$ of the same PC at $T$=4\,K.}
\label{fig5}
\end{figure}

The spectroscopic regime in PC can be confirmed by the presence of bosonic
features in $d^{2}V/dI^{2}$ spectrum. It is known that the
$d^{2}V/dI^{2}$ spectrum reflects the PC electron-phonon interaction function
$\alpha^{2}F(\omega )$ \cite{YansonZETF,Atlas,PCSbook}. The $d^{2}V/dI^{2}$ curves presented in Fig.\,\ref{fig5}
shows a clear maximum at about 20\,mV and a more smeared one around 60\,mV.
The measurements of the phonon density of states in isostructural
BaFe$_{2}$As$_{2}$ by inelastic electron scattering \cite{Mittal} showed a peak at
similar energies ($\approx $ 21.5\,meV) besides the more pronounced maxima at
13, 27 and 35\,meV. The latter maxima correspond well to the theoretical
calculation of the phonon density of state in BaFe$_{2}$As$_{2}$. The
authors claim, that the position of the peak at 21.5\,meV is not predicted by
the calculations, thus it may be due to electronic effects, e.g.,
Fermi-surface nesting. We should stress that mainly the backward scattering gives the
largest contribution to the PC spectra of electron-phonon interaction \cite{PCSbook}.
But, as it was shown in Ref.\cite{Kulic}, the backward scattering is substantially
suppressed in layered systems with strong electronic correlations. For the mentioned
reason, the observation of the phonon modes in PC spectra
of title compound can be complicated. Note, that the width of the peak at
20\,mV remains almost unchanged at the temperature rise. At the
same time, the phonon maxima should be smeared by the temperature increase,
but this is not the case. Thus, the phonon nature of this peak is doubtful.
However, we regularly measured such 20\,mV maximum for KFe$_{2}$As$_{2}$
samples \cite{Naidyuk3} which indicates that this feature is not an artifact and should
be intrinsic for [122]-type systems. Thus, additional investigations of that
feature are needed.

\section{Peculiarities in the superconducting state}
\begin{figure}[htbp]
\centerline{\includegraphics[width=9cm,angle=0]{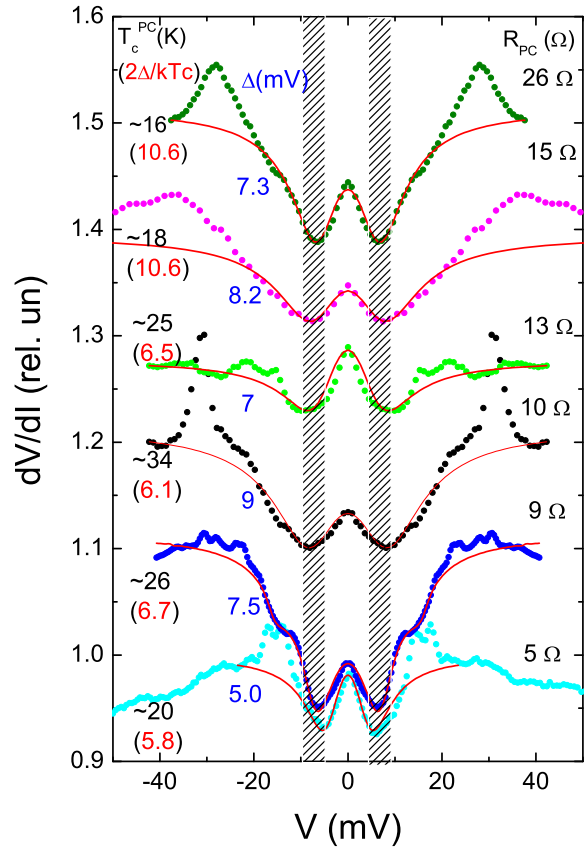}}
\caption{(Color online) Normalized on the normal state \textit{dV/dI(V)} data  (circles) displaying the AR-like features
(minima around $V$=0) for the PCs with different resistance $R_{PC}$ measured at 4\,K. The dashed vertical
stripes are placed to underline the position of the \textit{dV/dI(V)} minima which are
related to the SC gap. $T_{c}^{PC}$ is the approximate temperature at which SC
features disappear in the presented \textit{dV/dI(V).} The BTK fit (dashed curves) is shown for each curve 
with values of $\Delta $ and 2$\Delta /$k$_{\rm B}T_{c}$.  Two gap fit is
shown Fig.\,\ref{fig6} (see also Fig.\,\ref{fig8}). The curves are shifted for the sake of
clarity.}
\label{fig6}
\end{figure}

Let's turn to peculiarities of \textit{dV/dI(V)} appearing in the SC state, which are
related to the Andreev-reflection spectroscopy of the SC gap. It concerns
\textit{dV/dI} minima at energies corresponding to the SC energy gap, which are shown in
Fig.\,\ref{fig6} for several PCs with different resistance. Such \textit{dV/dI} characteristics correspond,
presumably, the spectroscopic (diffusive) regime of the current flow at least at low biases.
The dashed vertical stripes mark the position of the \textit{dV/dI} minima which appear
in the region between 6--9\,meV for different PCs.  Indeed, the gap value obtained from
ARPES measurements \cite{Aswartham} for the compound with maximal $T_{c}$ ($x=$0.4) is
around 10.5\,meV for the inner $\Gamma $ barrel and around
3\,meV for the outer $\Gamma $ barrel, while for the underdoped sample with
$x =$ 0.25 the estimated gap is about 2\,meV for the inner $\Gamma $ barrel.
So, the position of the minima in Fig.\,\ref{fig6} corresponds more to the gap value
for the compound with $x=$ 0.4, albeit our samples have $x=$ 0.25. It should be noted that the SC features in
\textit{dV/dI} presented in Fig.\,\ref{fig6} disappeared at considerably higher temperature than the bulk
$T_c\simeq$ 10\,K for compound with $x=$0.25, likely due to a higher $T_{c}$ of the surface
because of its inhomogeneity. That is, PCs have a larger local $T_{c}^{PC}$ than the bulk $T_{c}$.
\begin{figure}[htbp]
\centerline{\includegraphics[width=9cm,angle=0]{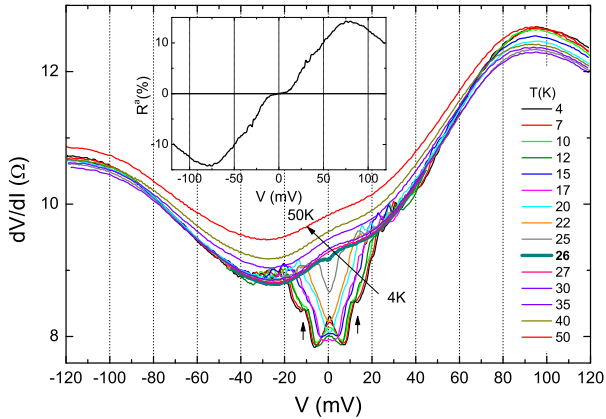}}
\caption{(Color online) \textit{dV/dI(V)} of the PC with $R\approx$ 9\,$\Omega $ which exhibits additional
minima seen at low temperatures (marked by the arrows). The
thick curve at 26\,K separates SC features (with minima) from normal state
behavior. Thus, $T=$ 26\,K can be taken as a local $T_{c}$. Inset: reduced
antisymmetric part $R^a$ calculated for curve at $T=$ 4\,K.}
\label{fig7}
\end{figure}

We fitted \textit{dV/dI(V)} in Fig.\,\ref{fig6} by widely accepted BTK model (see, e.g. Ref.\cite{Daghero})
to determine the gap value more precisely. As it is seen from Fig.\,\ref{fig6}, the gap value is
distributed between 5 and 9\,meV, that is it corresponds, in general, to the minima position in \textit{dV/dI(V)}.
At the same time the reduced gap 2$\Delta /$k$_{\rm B}T_{c}$ concentrated around 6 for 4 PCs from the bottom
and around 10 for two upper PCs if we take into account local $T_{c}^{PC}$. This correlates with the values reported
in \cite{Szabo} for Ba$_{1-x}$K$_{x}$Fe$_{2}$As$_{2}$, namely, 2$\Delta /$k$_{\rm B}T_{c}$=2.5--4 for a small
and 9--10 for a large gap.

\begin{figure}[htbp]
\centerline{\includegraphics[width=9cm,angle=0]{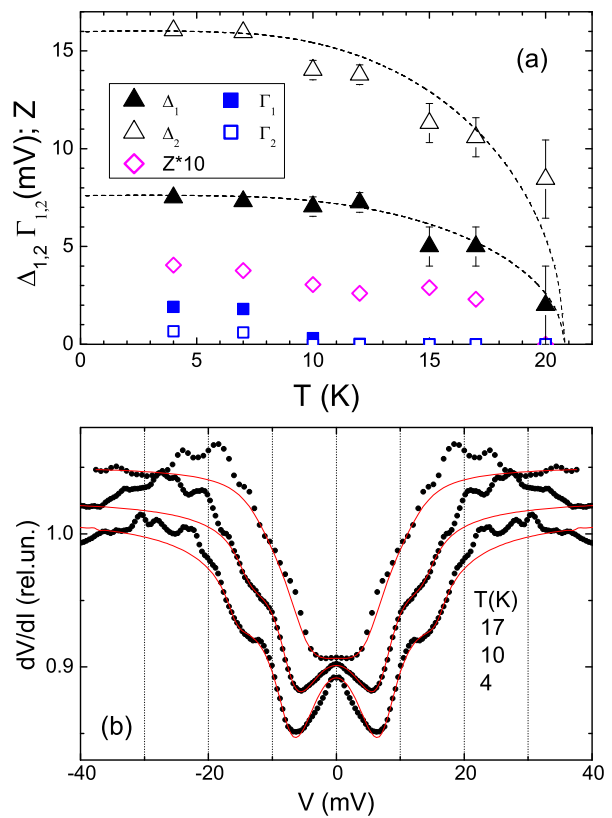}}
\caption{(Color online) (a) Temperature dependence of the SC gaps $\Delta $,
broadening parameters $\Gamma $, barrier strengths Z received from the BTK
fit in the two-gap approach for the PC from Fig.\,\ref{fig7}. The contribution of the
smaller gap turned out to be in the range 80-90{\%}. Dashed lines are BCS-like curves.
(b) Examples of the experimental \textit{dV/dI} (points) and calculated ones (lines)
within the two-gap fit curves for several temperature values ($T=$4, 10, and
17\,K).}
\label{fig8}
\end{figure}

We have measured a few \textit{dV/dI(V)} characteristics where the second pair of minima can
be connected, supposedly, to the second larger SC gap. The
temperature series for such PC is shown in Fig.\,\ref{fig7}. We have fitted these
\textit{dV/dI(V)} using the conventional two-gap approach \cite{Daghero} and found a satisfactory
agreement between the experimental and calculated curves with the SC gap
values $\Delta_{1}\approx $ 7.5\,meV and $\Delta_{2}\approx $
16\,meV at low temperatures (Fig.\,\ref{fig8}). Using the fit we have received the
temperature dependence of both SC gaps along with the utilized fitting
parameters, i.e. the barrier strength $Z$ and broadening parameters $\Gamma $.
However, the fit at higher temperatures becomes worse (especially by
approaching the local $T_{c}^{PC})$, what can be seen also from the
scattering of the fitting parameters in Fig.\,\ref{fig8}.
Also, the contribution of the large gap to the fit amounts to about 10{\%}
only, similar as for a few other PCs with resembling AR features.

The estimated values of 2$\Delta /$k$_{\rm B}T_{c}$ for this PC are turned out
to be quite high, about 8 and 18 for the small and the larger gap,
respectively, when we take $T_{c}\approx$21\,K from the extrapolation of the gap
dependence by the BCS curve (see Fig.\,\ref{fig8}). If we take the local $T_{c}\approx$26\,K
from Fig.\,\ref{fig7}, then the reduced 2$\Delta /$k$_{\rm B}T_{c}$ value will be $\approx$7 for the small gap
and still unrealistically large $\approx$15 for the large gap.
So, the features which we have taken as a larger gap are, apparently, of
different origin, while the much more reproducible minima around $\pm$ 6\,mV
are caused by the SC gap likely for the inner $\Gamma $ barrel.  If the bias
voltage further increases, \textit{dV/dI(V)} displays a broad maximum (Fig.\,\ref{fig7}) at about 90\,mV
and an asymmetric feature similar to that for PC in Fig.\,\ref{fig1}. Thus, the
contact region passes into the thermal regime. An additional evidence for
the nonspectroscopic origin of the second pair of minima is the behavior of
$R^{a}(V)$ for this PC shown in the inset of Fig.\,\ref{fig7}. At the energies of these
peculiarities, $R^{a}(V)$ starts deviate from zero what testifies that temperature in the PC core
increases above $T_{c}$. This raises doubts that the
mentioned features are due to the second SC gap.

It is also necessary to take into account that the larger gap should be more
suppressed and smeared by elastic scattering than the small one \cite{Efremov}.
Furthermore, the recent STM measurements on similar [122] compounds \cite{Shan,Song}
demonstrate that SC gap varies by factor two over scanned region of about 10\,nm.
The size of PC is at least a few times larger, as it was estimated
above. That is, we measure by PC some averaged gap or, it is not excluded,
merging of small and large gaps. Apparently, the more reproducible value of 2$\Delta /$k$_{\rm B}T_{c}$
of about 6, corresponds to such averaged gap. Interesting, that measured in \cite{Song} all
\textit{dI/dV} display only one pair of coherence maxima corresponding to a single gap.
Further, much broader coherence maxima from the larger gap can be
identified in \cite{Shan} only in some ``bright'' region. All this testify
that the large gap is more difficult to uncover.



\section{CONCLUSIONS}
Our PC study of iron-based superconductor Ba$_{1-x}$Na$_{x}$Fe$_{2}$As$_{2}$
($x\approx$ 0.25) shows that the current regime in PCs is at least diffusive at low
biases and it crosses to the thermal regime for increasing bias. In this
case, pronounced peculiarities in \textit{dV/dI(V)} data at high biases such as maxima and
asymmetry are caused by the specific resistivity $\rho (T)$ and the thermopower
$S(T)$ behavior versus temperature. The steep increase (or kink) in $\rho (T)$ due to
the antiferromagnetic transition produces \textit{dV/dI} maxima and thermoelectric effects
result in the asymmetry of \textit{dV/dI(V)}, which is well pronounced because of the
large thermopower in this system. We suppose that the thermal regime (or
heating) starts when antisymmetric part of \textit{dV/dI(V)} begins deviate from zero, what
can be used to distinguish between thermal and not thermal current flow in
PCs. We should also emphasize that an important selection rule for PC used
for investigation of normal state and SC properties of materials is the
similarity between $R_{PC}(T, V=0)$ and the bulk $\rho (T)$. A discrepancy between
them testifies that the substance in the PC core is disturbed and all
information taken from such PC spectra must be interpreted with
circumspection. Meanwhile, reproducible zero-bias minima detected in
\textit{dV/dI(V)} data in the range $\pm$(6--9)\,mV well below $T_{c}$ could be connected
with the manifestation of the SC gap. Features related to a second gap
turned out hard to resolve unambiguously, likely due to enhanced elastic
scattering, spatial inhomogeneity (reported in STM measurements) and transition to a thermal regime.
It is not excluded also merging of small and large gaps because of interband scattering caused by impurities.

\section*{Acknowledgements}
Yu.G.N. and O.E.K. thank V. Grinenko for assistance, the IFW Dresden for hospitality and the
Alexander von Humboldt Foundation for financial support in the frame of
a research group linkage program.

\section*{Appendix A}
\begin{figure}[htbp]
\centerline{\includegraphics[width=9cm,angle=0]{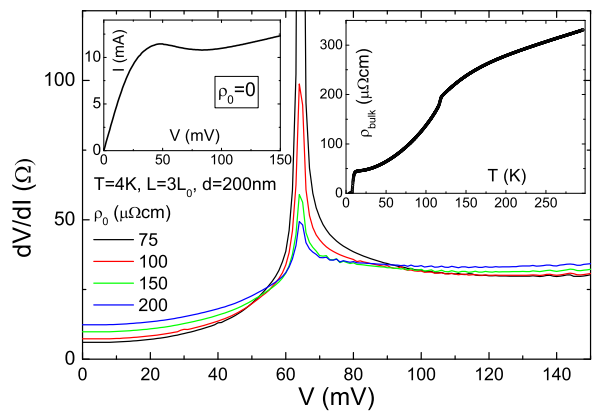}}
\caption{(Color online) \textit{dV/dI} curves calculated from Eqs.\,(\ref{eq2}),(\ref{eq3}) at 4\,K for several values of
residual resistivity using the bulk $\rho (T)$ of
Ba$_{0.75}$Na$_{0.25}$Fe$_{2}$As$_{2}$ shown in the right inset. Left
inset: calculated $I(V)$ curve for $\rho_{0}=$ 0.}
\label{fig9}
\end{figure}

In this Appendix we demonstrate the effect of the residual resistivity, temperature and
particular shape of $\rho (T)$ on the $dV/dI$ curves calculated with Eq.(3).
The calculated \textit{dV/dI} curve in Fig.\,\ref{fig2} shows a good
correlation with the experimental data, when we used experimentally corrected resistivity for PC. Here we performed a more detailed
calculation and analysis of PC \textit{dV/dI(V)} characteristics in the thermal regime
using also the bulk resistivity $\rho (T)$ for Ba$_{0.75}$Na$_{0.25}$Fe$_{2}$As$_{2}$ taken from Ref.\cite{Aswartham}.
The calculations are shown in Fig.\,\ref{fig9} for different values of residual resistivity $\rho_{0}$ at
4\,K. Our experimental \textit{dV/dI(V)} in Fig.\,\ref{fig2} have smaller amplitude and a broader maximum as
compared to the theoretical calculations. This can be due to a higher
residual resistance $\rho_{0}$ in the PC core caused by a non-perfect
structure at the interface, surface impurities, stress induced defects by PC
formation etc. As shown in Fig.\,\ref{fig9}, the specific value of the residual
resistivity does not affect the position of the maximum of \textit{dV/dI(V)}, however, the
maximum becomes smaller for increasing $\rho_{0}$. In the left inset, we
show the $I(V)$ curve for the case of $\rho_{0}=$ 0. It has the specific
$N$-like shape, where \textit{dV/dI(V)} diverges at the extrema of the $I(V)$ curve. We have not
observed such $I(V)$ in this study, but we have measured the $N$-like shaped $I(V)$ for
UPd$_{2}$Al$_{3}$ compound with similar behavior (steep increase) of $\rho (T)$ \cite{Naidyuk4}, what proves
the calculations.

\begin{figure}[htbp]
\centerline{\includegraphics[width=9cm,angle=0]{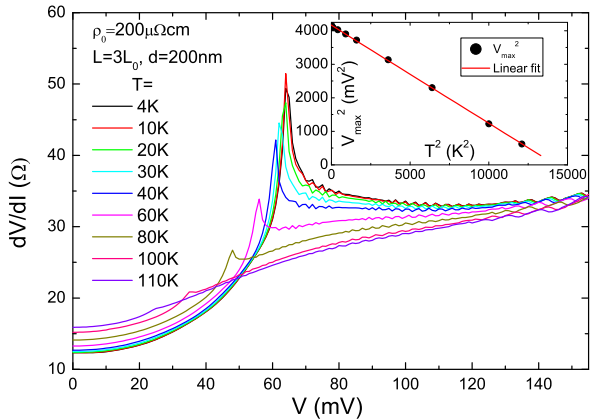}}
\caption{(Color online) \textit{dV/dI} curves calculated from Eqs.\,(\ref{eq2}),(\ref{eq3}) at several temperatures for $\rho
_0=200\,\mu \Omega$cm using bulk $\rho (T)$ of
Ba$_{0.75}$Na$_{0.25}$Fe$_{2}$As$_{2}$. Inset: position of maximum in
\textit{dV/dI} vs temperature in quadratic coordinates.}
\label{fig10}
\end{figure}

We have calculated the \textit{dV/dI(V)} for $\rho_{0}=$200$\,\mu \Omega $cm (as
estimated for the PC from Fig.\,\ref{fig1}) at different temperatures (Fig.\,\ref{fig10}). It
turns out that the position of the \textit{dV/dI(V)} maximum shifts to lower energies as the
temperature rises. Furthermore its position versus temperature follows a
quadratic law (see inset in Fig.\,\ref{fig10}) in accord with Eq.\,(2) supporting
our assumption of the thermal regime of the current flow through the PC.

\begin{figure}[htbp]
\centerline{\includegraphics[width=9cm,angle=0]{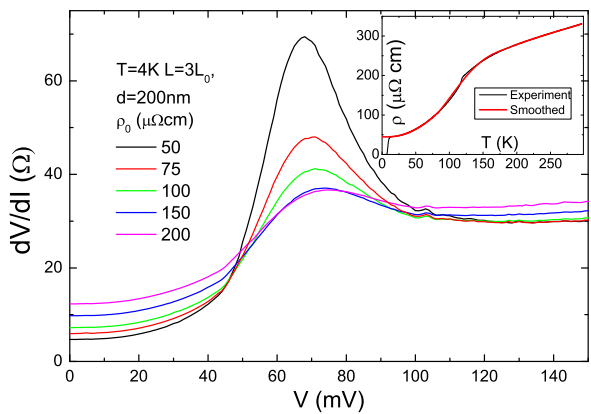}}
\caption{(Color online) \textit{dV/dI} curves calculated from Eqs.\,(\ref{eq2}),(\ref{eq3}) at 4\,K for several values of $\rho
_{0}$ using the smoothed $\rho (T)$ dependence (thick red curve) of
Ba$_{0.75}$Na$_{0.25}$Fe$_{2}$As$_{2\,}$ shown in the inset, where the thin black
curve shows original $\rho (T).$}
\label{fig11}
\end{figure}

It is naturally to suppose that not only the residual resistivity in the PC
is enhanced as compared to the bulk, but also $\rho (T)$ in PC core might be
modified. To model this situation, we slightly smoothed the $\rho (T)$ data
between $\sim$ 100\,K and 150\,K as shown in the inset of Fig.\,\ref{fig11}. The
smoothing of the sharp kink in $\rho (T)$ leads also to a smearing of
the corresponding maximum in \textit{dV/dI(V)} (see Fig.\,\ref{fig11}). Here we again present the
calculated \textit{dV/dI(V)} for different values of $\rho_{0}$ at $T=$4\,K using
the smoothed $\rho (T)$ dependence shown in the inset.

\begin{figure}[htbp]
\centerline{\includegraphics[width=9cm,angle=0]{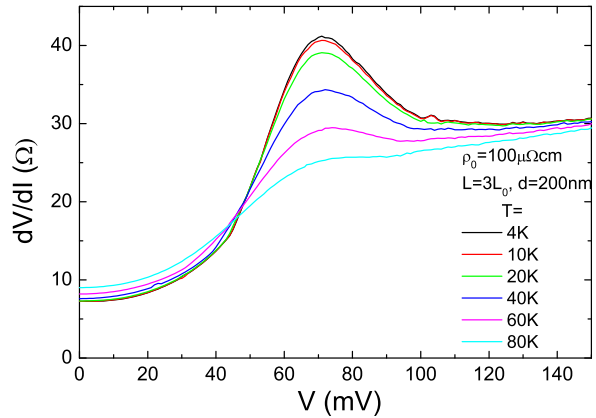}}
\caption{(Color online) \textit{dV/dI} curves calculated from Eqs.\,(\ref{eq2}),(\ref{eq3}) at several temperatures for $\rho
_{0}=$100 $\mu \Omega $cm using the smoothed $\rho (T)$ dependence from Fig.\,\ref{fig11} (inset).}
\label{fig12}
\end{figure}

In Fig.\,\ref{fig12}, the calculated temperature variation of \textit{dV/dI(V)} for $\rho_{0}=$100
$\mu \Omega $cm is presented. It is seen that the calculated
\textit{dV/dI(V)} curves have a similar shape as the measured \textit{dV/dI(V)} data shown in Fig.\,\ref{fig7}. In this
specific case, the maximum broadens and even slightly shifts to larger
voltages for increasing temperature, whereas in Fig.\,\ref{fig10} a shift of the sharp maximum
to smaller voltages is observed for increasing temperature. Thus, it depends
on the shape of the resistivity $\rho (T)$ in which direction the maximum of \textit{dV/dI} is shifted as function of the
temperature in the thermal regime of the current
flow. Only, if the maximum in \textit{dV/dI} is quite sharp, its position shifts according to Eq.\,(\ref{eq2}).

\begin{figure}[htbp]
\centerline{\includegraphics[width=9cm,angle=0]{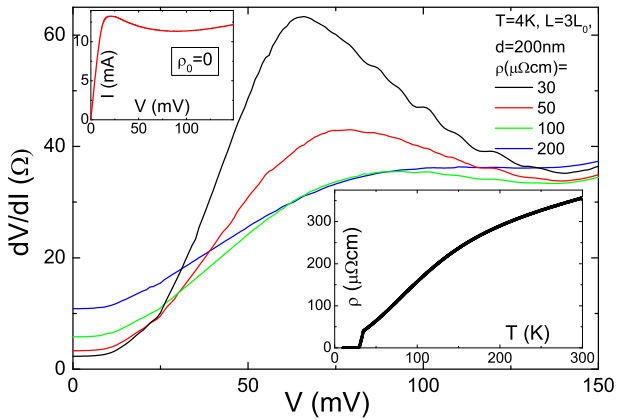}}
\caption{(Color online) \textit{dV/dI} curves calculated from Eqs.\,(\ref{eq2}),(\ref{eq3}) at 4\,K for several values of
$\rho_{0}$ using $\rho (T)$ of Ba$_{0.65}$Na$_{0.35}$Fe$_{2}$As$_{2}$ \cite{Aswartham}
shown in the right inset. Left inset: calculated $I(V)$ curve for $\rho
_{0}=$ 0.}
\label{fig13}
\end{figure}

\begin{figure}[htbp]
\centerline{\includegraphics[width=9cm,angle=0]{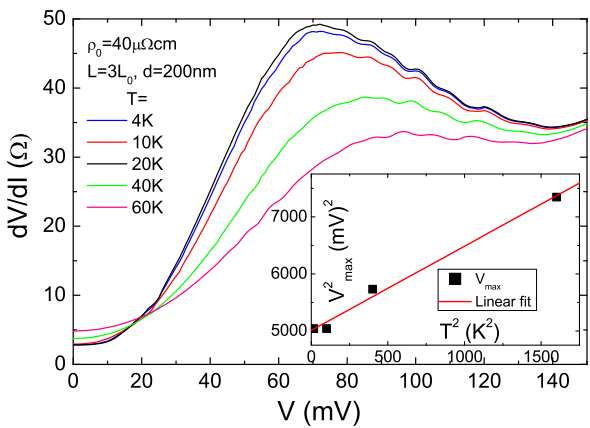}}
\caption{(Color online) \textit{dV/dI} curves calculated from Eqs.\,(\ref{eq2}),(\ref{eq3}) at several temperatures for
$\rho_{0}=40\,\mu \Omega $cm using $\rho (T)$ of
Ba$_{0.65}$Na$_{0.35}$Fe$_{2}$As$_{2}$. Inset: the position of maximum in
\textit{dV/dI} versus temperature in the quadratic coordinates.}
\label{fig14}
\end{figure}

Finally, due to enhanced local  $T_{c}$ observed for many PCs, we have also
calculated \textit{dV/dI(V)} from $\rho (T)$ for a compound with $x=$ 0.35 and $T_{c}=$ 30\,K. Fig.\,\ref{fig13}
 presents the calculated \textit{dV/dI(V)} for different $\rho_{0\, }$based on $\rho
(T)$ of this compound (see right inset in Fig.\,\ref{fig13}). The $I(V)$ characteristic for
$\rho_{0}=$0 has again the N-shape, while for increasing $\rho_{0}$
a broadening and shifting of the \textit{dV/dI} maximum to larger voltages is found. The
temperature variation of \textit{dV/dI(V)} for $\rho_{0}=$ 40\,$\mu \Omega$cm plotted
in Fig.\,\ref{fig14} shows a broadening and shift of the \textit{dV/dI(V)} maximum to larger voltages
for increasing temperature. The position of the maximum in \textit{dV/dI(V)} versus
temperature follows approximately a quadratic law, but opposite to that in inset in Fig.\,\ref{fig10}.

Summarizing our calculations, we should note that:
\begin{itemize}
\item We can reproduce the measured \textit{dV/dI(V)} supposing the realization of the thermal regime
in PCs and some modification of $\rho (T)$ in the contact core.
\item The position of the peculiarities (peaks) of \textit{dV/dI(V)} characteristics in the thermal regime can be
shifted both to lower voltages (for sharp peaks) as well as to higher voltages (for broad maxima) as the temperature rises.
\item The variation of the peak position with temperatures can be described by Eq.\,(\ref{eq2}) only if the peculiarities
are quite sharp.
\end{itemize}

\section*{Appendix B}

\begin{figure}[htbp]
\centerline{\includegraphics[width=9cm,angle=0]{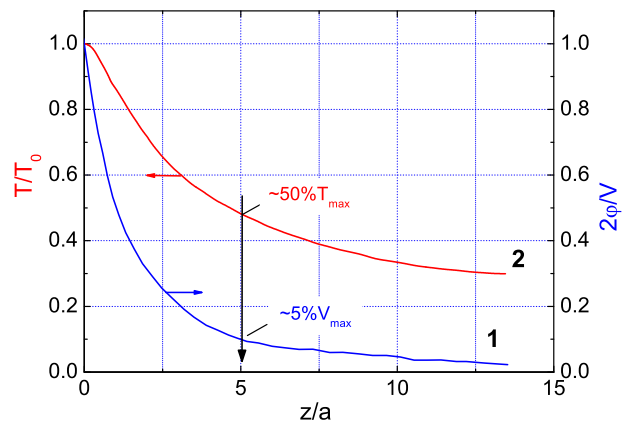}}
\caption{(Color online) Distribution of the potential (1) and temperature (2) from the PC
center $z$=0 along the $z$-axis perpendicular to the plane of PC in the
thermal regime after Kulik theory \cite{Kulik2}. Long vertical arrow marks the
residual potential and temperature values at $z=5a$, where $a=d/2$ is the PC
radius.}
\label{fig15}
\end{figure}

We should note that \textit{dV/dI(V)} remain asymmetric in the SC state below $T_{c}$ (see,
e.g., the data at 4.5\,K in Fig.\,\ref{fig1}). However, it is known that the thermopower
of any conductor equals to zero in the SC state. That is why the
investigated superconductor in the contact region should be in the normal
state at biases larger than the energy of SC peculiarities (gaps) of
\textit{dV/dI(V)}. It takes place because of heating and transition of the PC core in the normal
state at a voltage increase. Next, as it was shown by Kulik \cite{Kulik2}, in the thermal regime the
temperature within the PC decreases gradually (from its maximum value in the plane of
the PC) along the $z$-axis perpendicular to the plane (see Fig.\,\ref{fig15}). At the same time, as
Fig.\,\ref{fig15} shows, the potential drops much faster from the PC center than the
temperature decreases. Thus, at a distance $\approx$\,2.5$d$ from the PC
center the potential deviates only on a few percents from its equilibrium,
while the temperature at this distance still amounts a half from that in the
PC center (see Fig.\,\ref{fig15}). Moreover, for heterocontacts in the thermal limit
the maximal temperature is achieved in the metal with the larger resistivity
and not in the PC center \cite{Itskovich}. Thus, due to heating, "hot spot" or the normal state region
spreads effectively further from PC core (where main potential gradient is concentrated) and it has properties more
close to the bulk material, what is reflected in a more reproducible shape
of the antisymmetric part of the \textit{dV/dI} of the PC in Fig.\,\ref{fig4} with different raw \textit{dV/dI }behavior.


\end{document}